\documentstyle[aps,epsf]{revtex}  

\begin{document}        

\baselineskip 14pt
\title{Structure Functions, the Gluon Density, and PQCD Tests}
\author{Richard Cross}
\address{University of Wisconsin-Madison}
\address{For the ZEUS collaboration}
\maketitle              

\begin{abstract}        
Measurements of the proton structure function $F_2$ for
$0.11<Q^2<20000$ $GeV^2$ and $1.2\times 10^{-5} < x < 0.65$
from ZEUS 1994-1997 measurements are presented.
From ZEUS 1994 and 1995 $F_2$ data the slopes $dF_2/d\ln Q^2$ at fixed $x$
and $d\ln F_2/d\ln(1/x)$ for $x<0.01$ at fixed $Q^2$ are derived.
For the latter E665 data are also used. The transition region in $Q^2$
is explored using the simplest non-perturbative models and NLO QCD.
The data at very low $Q^2$ $\leq 0.65 GeV^2$ are described
successfully by 
Regge theory. From a NLO QCD fit to ZEUS data the gluon density in the
proton is extracted in the range $3\times 10^{-5}<x<0.7$.
Data from NMC and BCDMS constrain the fit at large $x$.
Assuming the NLO QCD description to be valid down to $Q^2\sim 1 GeV^2$,
it is found that the $q\bar{q}$ sea distribution is still
rising at small $x$ and the lowest $Q^2$ values whereas the gluon
distribution is strongly suppressed.  Preliminary ZEUS 1996-1997 $F_2$ 
measurements are also presented.
\
\end{abstract}          

\section{Introduction}               
\begin{figure}[ht]      
\centerline{\epsfxsize 2.65 truein \epsfbox{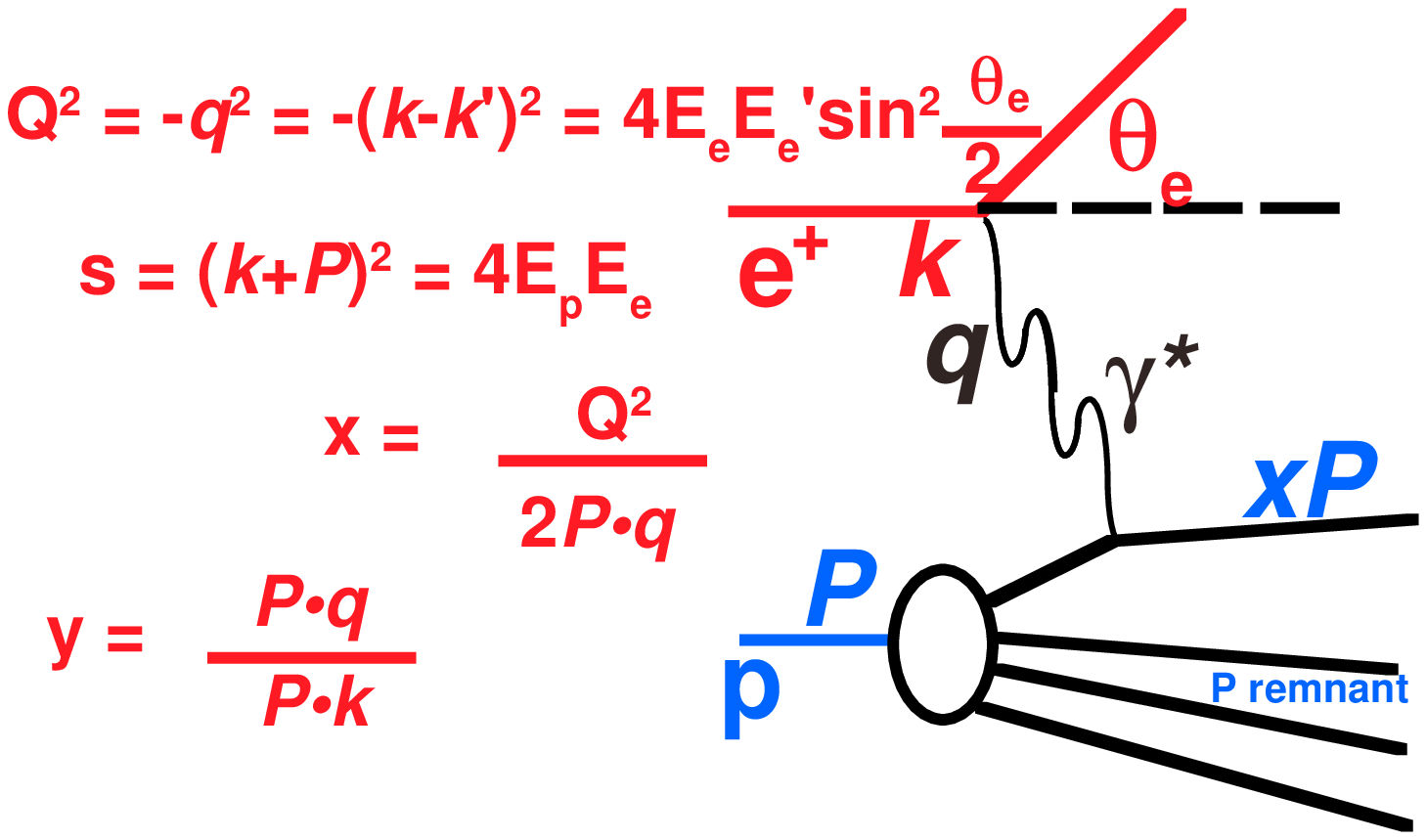}}
\vskip -.2 cm
\caption[]{
\label{figure1}
\small }
\end{figure}

The deep inelastic scattering process is used to probe the structure of
nucleons.  Ultra-high energy transfers are achieved at the DESY laboratory, 
in Hamburg, Germany, where 820 GeV protons are collided 
with 27.5 GeV positrons for a center-of-mass energy $\sqrt{s}$=300 GeV.
The deep inelastic scattering (DIS) process at HERA is illustrated in 
Fig. ~\ref{figure1}.
The variables are related by $Q^2 = sxy$.

In the $Q^2$ range examined here the double differential cross-section for 
single virtual-photon exchange in DIS is given by
\begin{equation}
  \frac{d^2\sigma}{dxdQ^2} = 
\frac{2\pi\alpha^2}{x Q^4}
\left[ 
2\left( 1-y \right) + \frac{y^2}{1+R}
\right] 
F_2(x,Q^2) 
\left[ 1 + \delta_r(x,Q^2) \right],
\label{eqn:d2dxdq2}
\end{equation}
where $R$ is related to the longitudinal structure function $F_L$ by 
$R=F_L/(F_2-F_L)$ and $\delta_r$ gives the radiative corrections to the 
Born cross-section, which in this kinematic range are at most $10\%$. 
For $R$ we take values given by the BKS model \cite{BKS}.

Measurements of $F_2$ for $0.11<Q^2<20000$ $\rm{GeV}^2$ are presented.
The rapid rise of $F_2$ at low $x$ can be seen.  The $F_2$
measurements are fit to pQCD and Regge motivated functions.  
These measurements 
illuminate the region where QCD breaks down because $Q^2 \rightarrow 0$.

The ZEUS 1995 $F_2$ measurement increased the kinematic coverage in the
low $x$ and $Q^2$ region.
The coverage for $Q^2$ between $0.11$ and $0.65$  $\rm{GeV}^2$ was made
possible by the installation of a small electromagnetic sampling
calorimeter, the Beam Pipe Calorimeter (BPC), at small positron
scattering angles.  The proton structure
function $F_2$ and the total $\gamma^*p$ cross-section 
have been published \cite{ZBPC}.
Results for $Q^2$ between $0.6$ and $17$ $\rm{GeV}^2$ were made possible
by a set of runs in 1995 with a mean interaction vertex shifted 70 cm 
in the direction of the incoming electron.  
The QCD fits presented also used data from the 1994 ZEUS $F_2$ measurement 
and from E665.

These data sets are fit using QCD and Regge functional forms.
The slopes are extracted.  The combined sample is fit using DGLAP. 
The results of the DGLAP fit are used to extract the gluon density.

Also shown are preliminary results the 1996-97 ZEUS $F_2$ measurements.
There is reasonable agreement with the 1994 ZEUS results.
The 1996-97 measurement fills in the region between the 1994 ZEUS results 
and results from fixed target experiments.
The 1996-97 results also have data at higher $Q^2$.

\section{The ZEUS detector.}

ZEUS~\cite{ref:zeusdet} is a multi-purpose magnetic detector.
The primary components used in these analyses are the uranium calorimeter, 
the beam pipe calorimeter, tracking detectors, and luminosity monitors.
The coordinate system is defined such that
the $z$-axis follows the proton direction, and the origin is
the nominal $ep$ interaction point.
 
The ZEUS compensating uranium-scintillator calorimeter covers the polar 
angle region 
$2.6^{\circ} < \theta < 176.1^{\circ}$ with full azimuthal coverage
over this region.
Its energy resolution for electromagnetic showers is 
$\sigma_E/E \simeq 18 \% / \sqrt{E(\rm{GeV})}$, and 
for hadronic showers is
$\sigma_E/E \simeq 35 \% / \sqrt{E(\rm{GeV})}$.
The Beam Pipe Calorimeter is placed outside the main Uranium calorimeter
and at an angle of almost $180^{\circ}$ so that
it can measure outgoing electrons at $Q^2$ down to $0.1$ $\rm{GeV}^2$.
The ZEUS tracking detectors primarily provide vertex reconstruction.
ZEUS has cylindrical central tracking chambers
and forward and rear tracking chambers, operating in a solenoidal
1.43 T magnetic field.
 
The luminosity is determined from the
rate of Bethe-Heitler bremsstrahlung ($ep\rightarrow ep\gamma$) events
in detectors near the beamline.

\section{$F_2$ fits and the transition region.}

Three ZEUS $F_2$ data samples are used to examine the transition region.
The ZEUS 1994 DIS ZEUS sample required an electron of 10 GeV in the uranium
calorimeter, $35$ GeV $< E-p_z < 65$ GeV,
and an event vertex in the range 
$40 \rm{cm} < Z_{\rm vertex} < 160 \rm{cm}$.  The 1995 shifted
vertex sample used a different nominal vertex, so the event vertex is required
to lie in the range $40 \rm{cm} < Z_{\rm vertex} < 160 \rm{cm}$.
The 1995 Beam Pipe Calorimeter sample required that an electron be
contained in the BPC rather than the uranium calorimeter.  Because the
BPC is at smaller $\eta$ than the uranium calorimeter, the BPC sample can reach
lower $Q^2$.

Low $Q^2$($Q^2 < 10 \rm{GeV}^2$) $F_2$ results at ZEUS show the transition region from pQCD to the Regge region.
1995 BPC data are binned in $W^2$ and fit to a Donnachie-Landshoff\cite{dltwo} 
Regge parameterization of the form 
\begin{displaymath}
\sigma_{\rm tot}^{\gamma p}(W^2) = A_R (W^2)^{\alpha_R-1}+ 
A_P(W^2)^{\alpha_P-1}
\end{displaymath}
where $P$ and $R$ denote the Pomeron and Reggeon contributions.  This is
the ZEUSREGGE curve in figure ~\ref{figure2}.

Results from the 1994 $F_2$, 1995 shifted vertex and E665 are
also fit with the QCD based model.  The kinematic range covered by this
data is $3 \times 10^{-5} < x < 0.7$ and $0.9 < Q^2 < 5000$ $\rm{GeV}^2$.
This data is fit by solving the DGLAP evolution equations at NLO in the 
{\mbox{$\overline{\rm{MS}}$} scheme~\cite{mbb:furm}.

\begin{figure}[ht]      
\centerline{\epsfxsize 2.5 truein \epsfbox{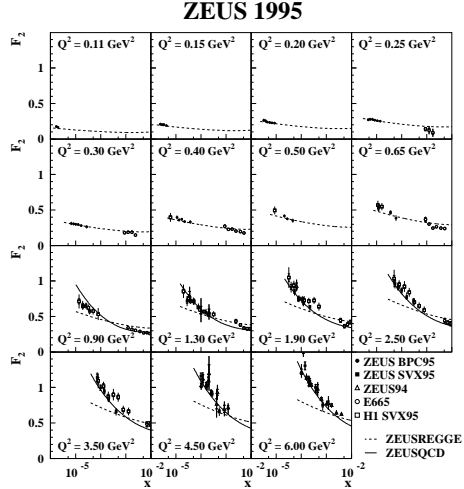}}
\vskip -.2 cm
\caption[]{
\label{figure2}
\small Low $Q^2$ $F_2$ data for different $Q^2$ bins.  A Regge inspired
fit(dashed lines) to the BPC95 data is shown.  At larger $Q^2$ values 
a NLO QCD fit to the ZEUS data is also shown.}
\end{figure}

The 1994, 1995 and E665 samples are binned in $Q^2$ and their slopes were 
extracted.  These slopes are compared with QCD and Regge predictions.  
A GRV parameterization and the 
ZEUS QCD fit are shown for the higher $Q^2$ data.  For the low $Q^2$ region
the Regge fit to ZEUS data and a global Donnachie-Landshoff\cite{dltwo}
fit are shown.
Neither type of parameterization can describe the data in the transition 
region from high to low $Q^2$.

\begin{figure}[ht]      
\centerline{\epsfxsize 3.0 truein \epsfbox{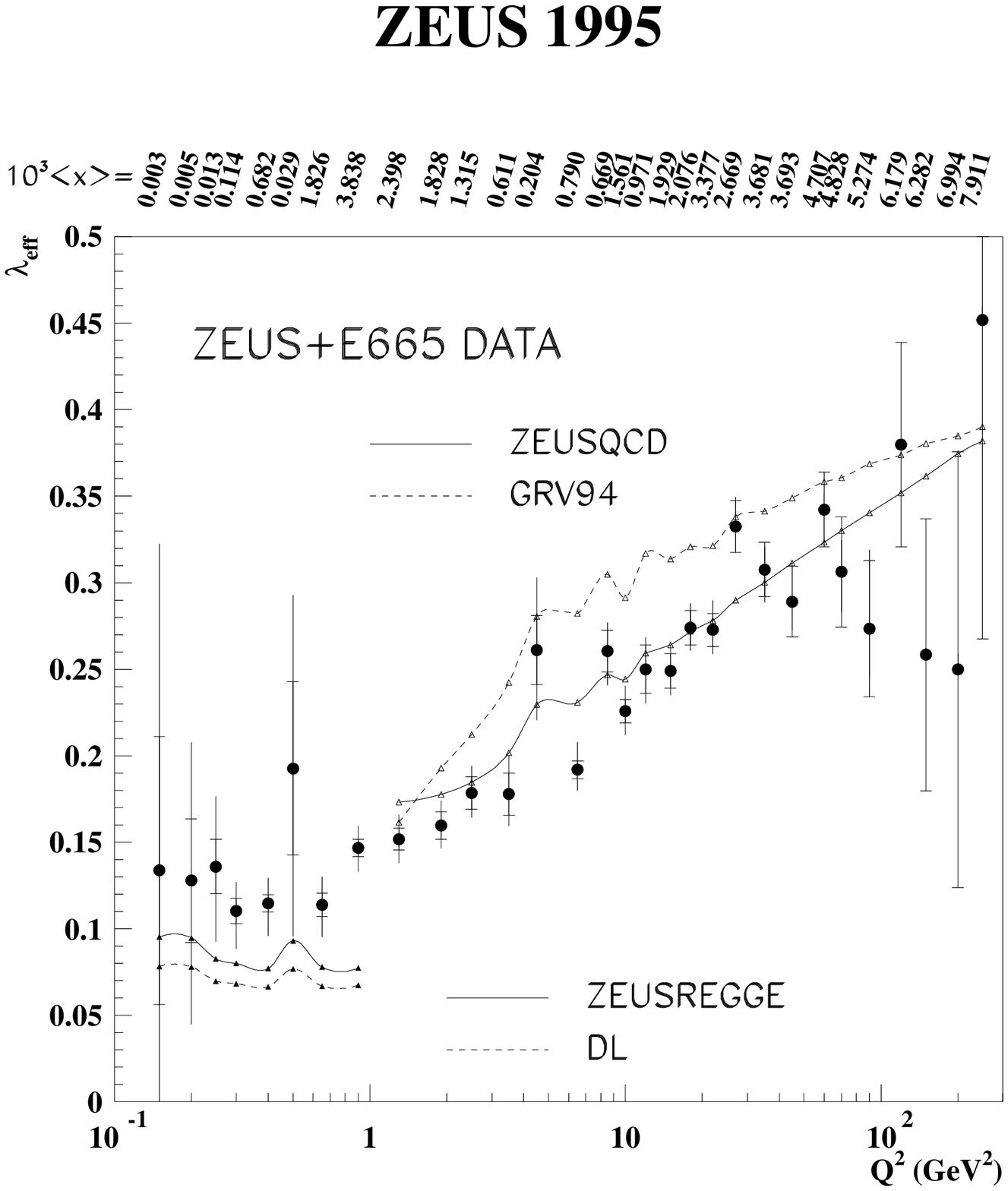}\epsfxsize 3.0 truein \epsfbox{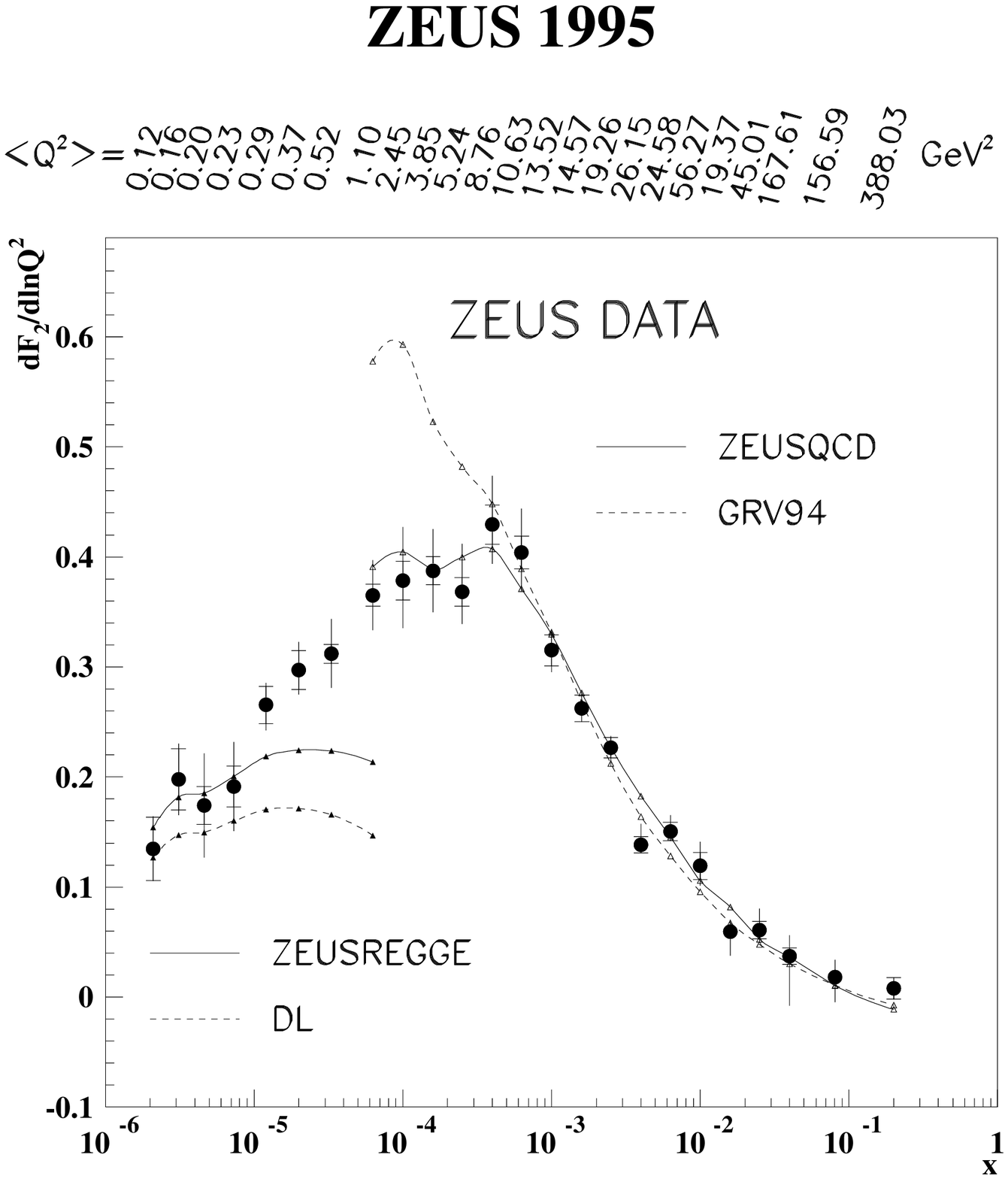}}   
\vskip -.2 cm
\caption[]{
\label{figure3}
\small 
$d\ln F_2/d\ln(1/x)$ as a function of $Q^2$ calculated
by fitting ZEUS and E665 $F_2$ data in bins of $Q^2$ to the functional form
$A x^{-\lambda_{eff}}$(left). 
$d F_2/d\ln Q^2$ as a function of $x$ calculated by fitting
ZEUS $F_2$ data in bins of $x$ to the functional form
$a+b\ln Q^2$(right). The inner error bar
shows the statistical error and the outer the total statistical
and systematic error added in quadrature. In a $Q^2$ bin 
$\langle x \rangle$ is calculated from the weighted mean of 
ln$x$. The linked points labelled DL 
and GRV94 are from the Donnachie-Landshoff Regge 
fit \cite{dltwo} and the GRV94 NLO QCD fit 
\cite{grv94}. In both cases the points are 
obtained using the same weighted range of $x$ as for the experimental data.
}
\end{figure}
The ZEUS94\cite{ZF2} and the 1995 shifted vertex data are fit  
in order to extract the gluon and quark densities at low $x$.
They are fit by solving the DGLAP~\cite{dglap} evolution equations in NLO 
in the {\mbox{$\overline{\rm{MS}}$} scheme~\cite{mbb:furm}.
For this fit the starting scale is $Q^2_0 = 7$ $\rm{GeV}^2$.  
For data with $Q^2 < 7$ $\rm{GeV}^2$, backwards evolution in $Q^2$ 
are performed.
The gluon distribution ($xg$), the sea quark distribution ($xS$) 
and the difference of up and down quarks in the proton ($x\Delta_{ud}$) are
parameterised as
\begin{eqnarray} \label{mbb:param}
xg(x,Q_0^2) & = & A_g x^{\delta_g}(1-x)^{\eta_g} 
(1 + \gamma_g x) \nonumber \\
xS(x,Q_0^2) & \equiv &  2x ( \bar{u} + \bar{d} + \bar{s} ) =
A_s x^{\delta_s}(1-x)^{\eta_s}(1+\varepsilon_s \sqrt{x}
 + \gamma_s x) \\
x\Delta_{ud}(x,Q_0^2) & \equiv & x(u+\bar{u}) - x(d+\bar{d}) =
A_{\Delta} x^{\delta_{\Delta}}(1-x)^{\eta_{\Delta}}.
  \nonumber 
\end{eqnarray}
The input valence distributions $xu_v = x(u-\bar{u})$ and 
$xd_v = x(d-\bar{d})$
at $Q^2_0$ are taken from the parton distribution set
MRS(R2)~\cite{mrsr}. As for MRS(R2) we assume that the 
strange quark distribution is a given fraction $K_s = 0.2$ of the sea
at the scale $Q^2 = 1 \rm{GeV}^2$. 
The gluon normalization, $A_g$, is fixed by the
momentum sum rule, that is, the total momentum fraction carried by quarks
and gluons is required to add up to unity. There are 11 free
parameters in the fit.

The fit yields a gluon distribution that rises dramatically as $x$ decreases
for $Q^2$ of $20 \rm{GeV}^2$.  The gluon distribution is consistent with 0 for 
$Q^2$ of $1 \rm{GeV}^2$.  
It is not constrained to be greater than 0 in the fit.
A negative gluon distribution is not in contradiction with NLO QCD as long
as the cross sections calculated from the parton distributions are
positive for all $x$ and $Q^2$ in the fitted kinematic region.  This is the
case for the fit values shown here.  The sea quark distribution rises at low
$x$ even for $Q^2$ of $\rm{GeV}^2$.  This can be interpreted as higher energy
gluons generating $q\bar{q}$ pairs.

\begin{figure}[ht]      
\centerline{\epsfxsize 3.5 truein \epsfbox{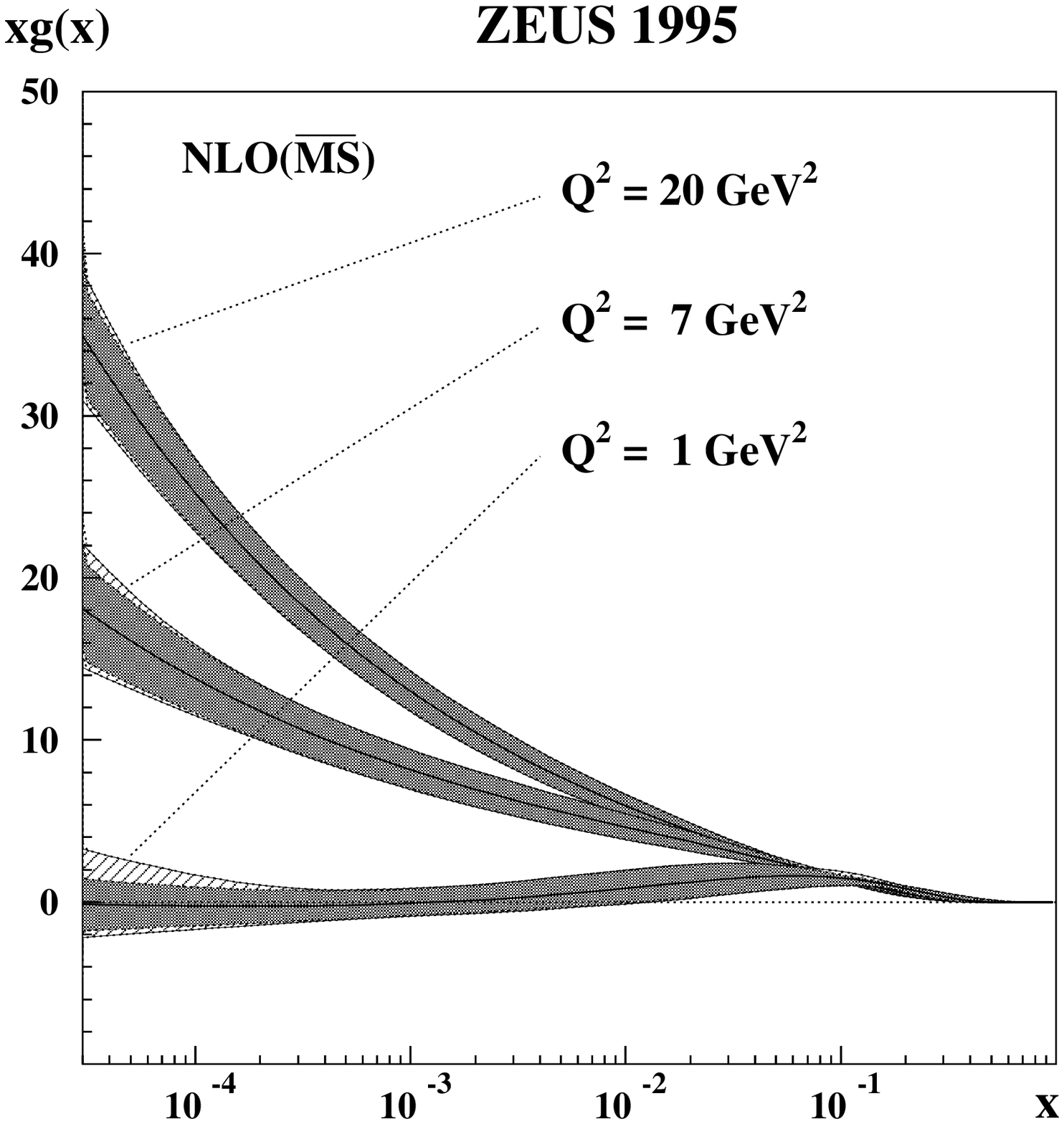}
\epsfxsize 3.5 truein \epsfbox{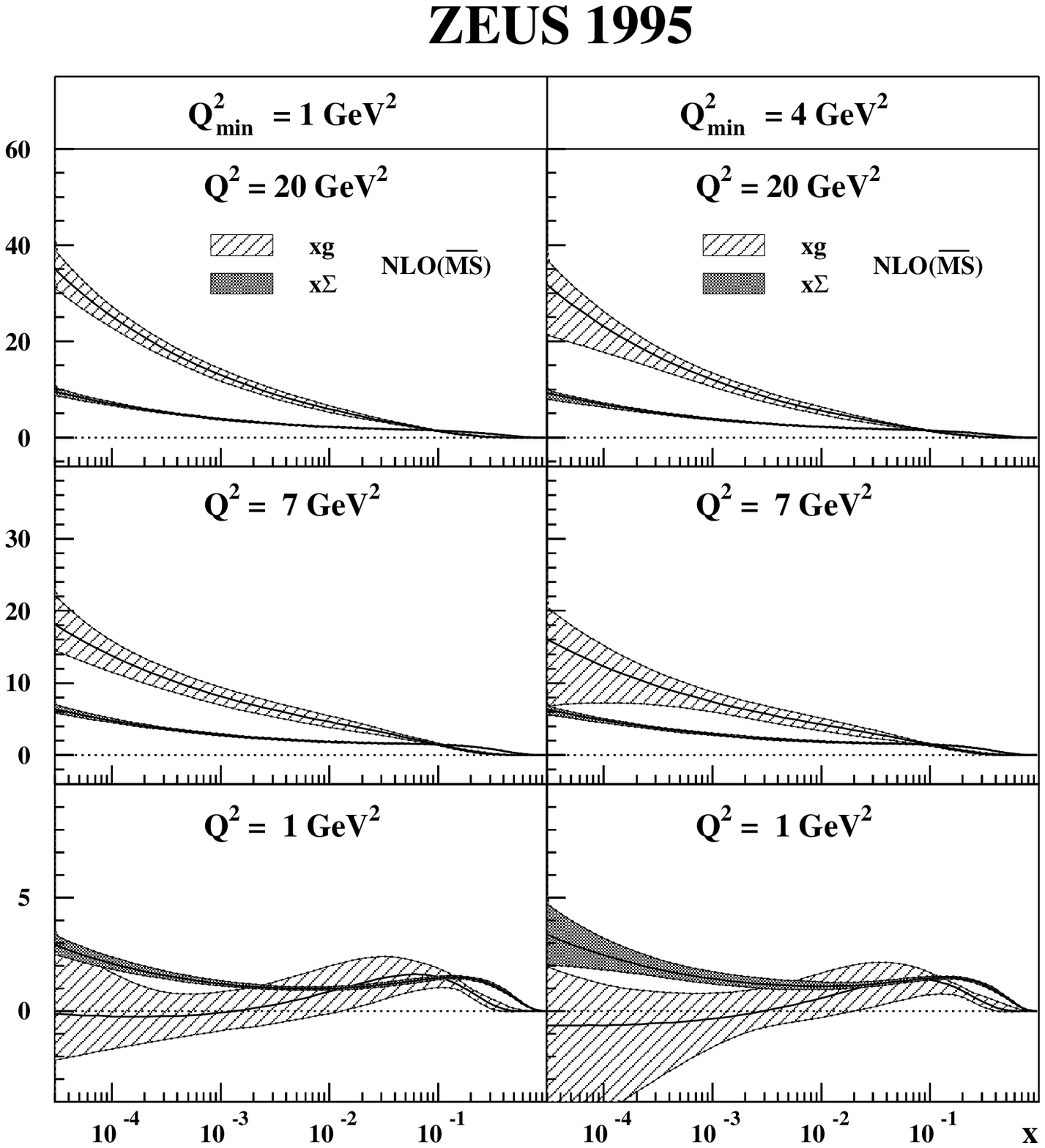}}   
\vskip -.2 cm
\caption[]{
\label{figure6}
\small The gluon distribution $xg(x)$ as a function of $x$(left).
The quark singlet momentum distribution, $x\Sigma$ (shaded), and 
the gluon momentum distribution, $xg(x)$ (hatched), as 
functions of $x$ at fixed
values of $Q^2 = 1$, 7 and 20 GeV(right). The error bands correspond to the
quadratic sum of all error sources considered for each parton density.
}
\end{figure}

\section{1996-97 $F_2$.}

The ZEUS 1996-1997 $F_2$ measurement is split into two samples.
The high $Q^2$ measurement uses the full $27.4 pb^{-1}$ of data taken in
1996-1997 run.  The low $Q^2$ measurement uses only $6.8 pb^{-1}$ of data.  
The combined measurements cover a $Q^2$ range $1.5 - 20000 \rm{GeV}^2$
and an x range of $2.6\times 10^{-5}<x<0.65$.  Most bins have systematic 
errors of 3-4\%.
At high and low y, the error is as high as 10\%.  The high $Q^2$ measurement 
has better statistics than the 1994 ZEUS $F_2$ measurement.

The data was taken by requiring a positron with energy $E_e > 10$ GeV.  
The accepted events were also required to have $44~$ GeV$ < E-p_z < 68$ GeV,
and an event vertex in the range $-50$ cm $< z < 50$ cm.

The efficiency of the cuts is determined by comparing the data to a
Monte Carlo that used the CTEQ4D parton densities. 
Figure ~\ref{figure7} shows a comparison of several data and Monte Carlo 
quantities.  The efficiency of the cuts as determined by Monte Carlo is 90\%.

\begin{figure}[ht]      
\centerline{\epsfxsize 3.5 truein \epsfbox{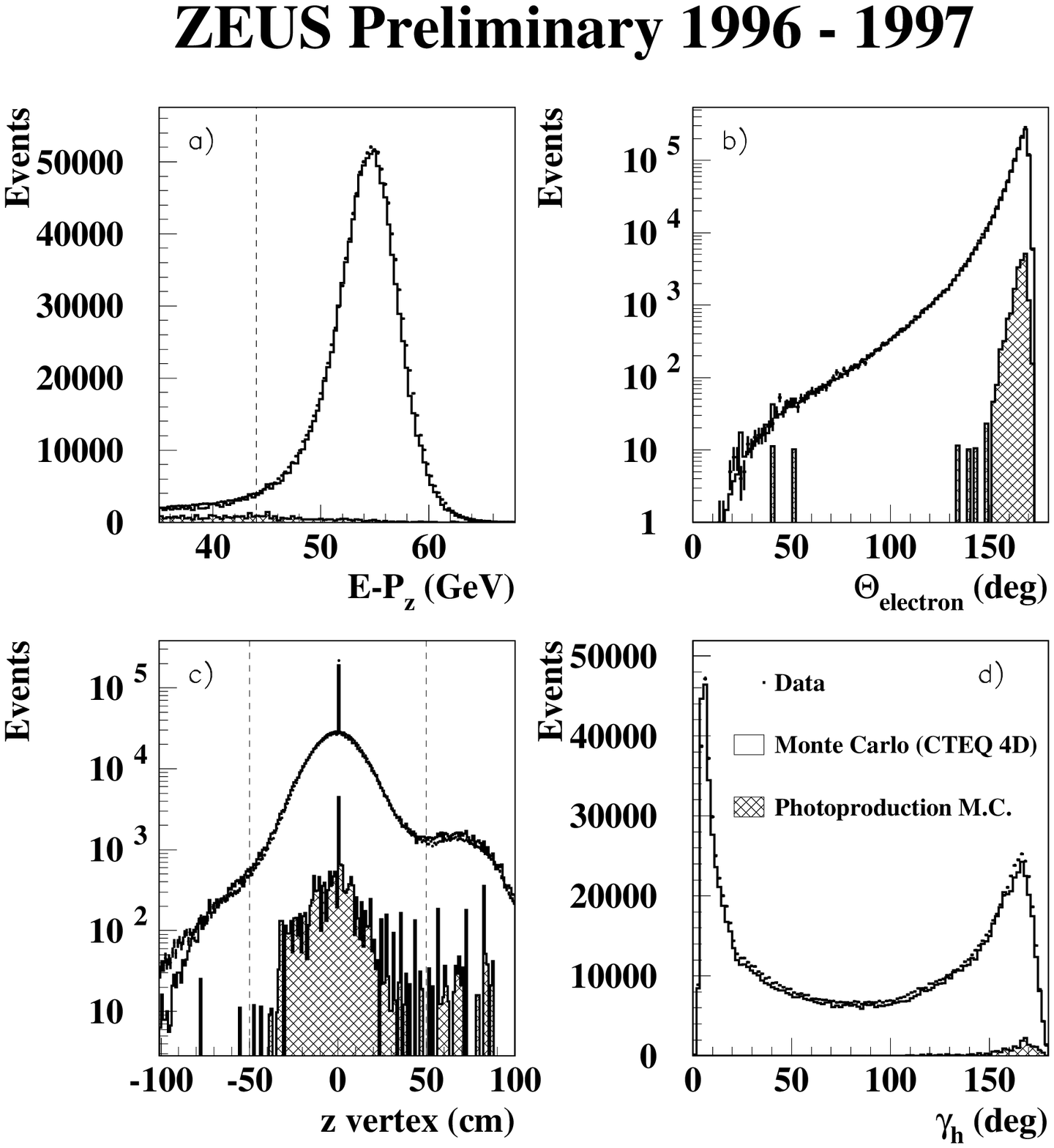}}   
\vskip -.2 cm
\caption[]{
\label{figure7}
\small a) $E - P_z$, b)  angle of the measured electron,
c) the z position of the event vertex, and d) the angle of the 
hadronic system.  The Monte Carlo(histograms) are normailised to
the luminosity of the data(dots).  Data and MC show good agreement.
The cuts requiring energy greater than 45 GeV (a) and an event vertex
between + 50 cm in z and - 50 cm in z (c) are shown.  
}
\end{figure}

The ZEUS 1996-1997 $F_2$ data sample fills in many of the gaps between fixed
target experiments and the 1994 ZEUS $F_2$ sample.  The errors are smaller than
the 1994 sample.  The 1996-1997 sample shows agreement with the 1994 sample
and with previous fixed target experiments.

For a fixed $Q^2$, the ZEUS 1996-1997 $F_2$ rises dramatically as $x$
decreases(see Figures ~\ref{figure8} and ~\ref{figure9}).  Also, for a 
fixed $x$, $F_2$ rises with $Q^2$(see Figure ~\ref{figure10}).

\begin{figure}[ht]      
\centerline{\epsfxsize 3.5 truein \epsfbox{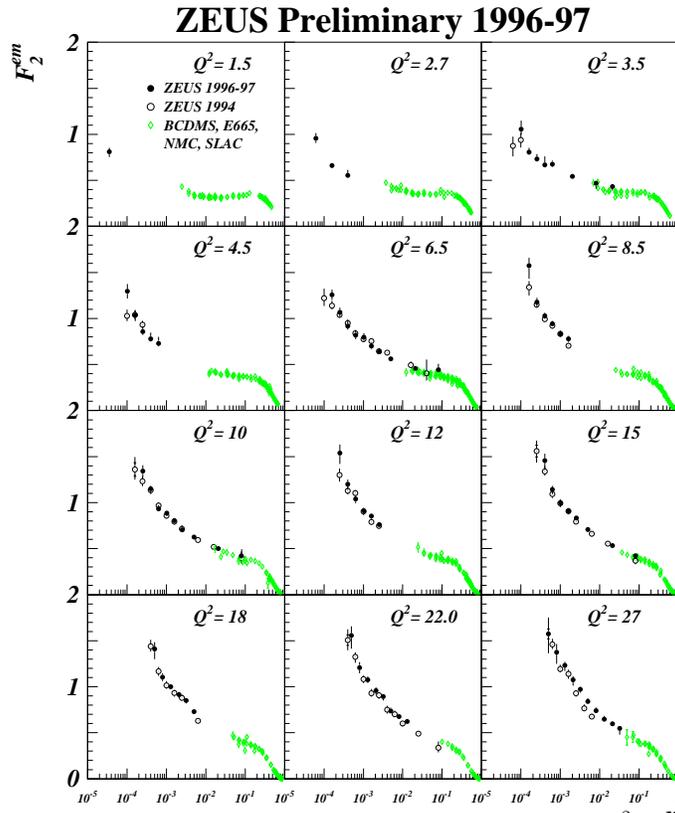}}   
\vskip -.2 cm
\caption[]{
\label{figure8}
\small ZEUS 1996 1997 $F_2$ measurement at low $Q^2$.}
\end{figure}

\begin{figure}[ht]      
\centerline{\epsfxsize 3.5 truein \epsfbox{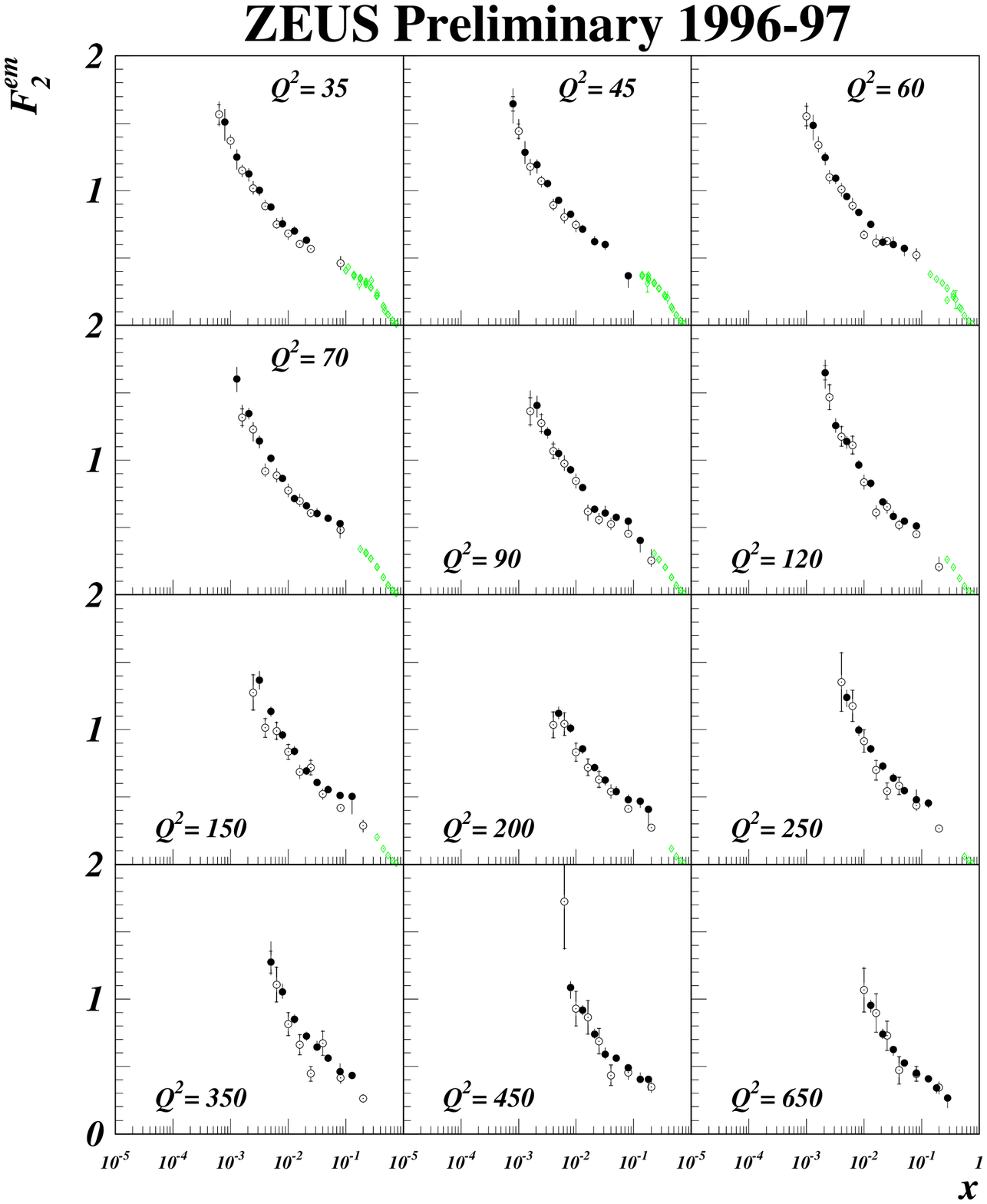}\epsfxsize 3.5 truein \epsfbox{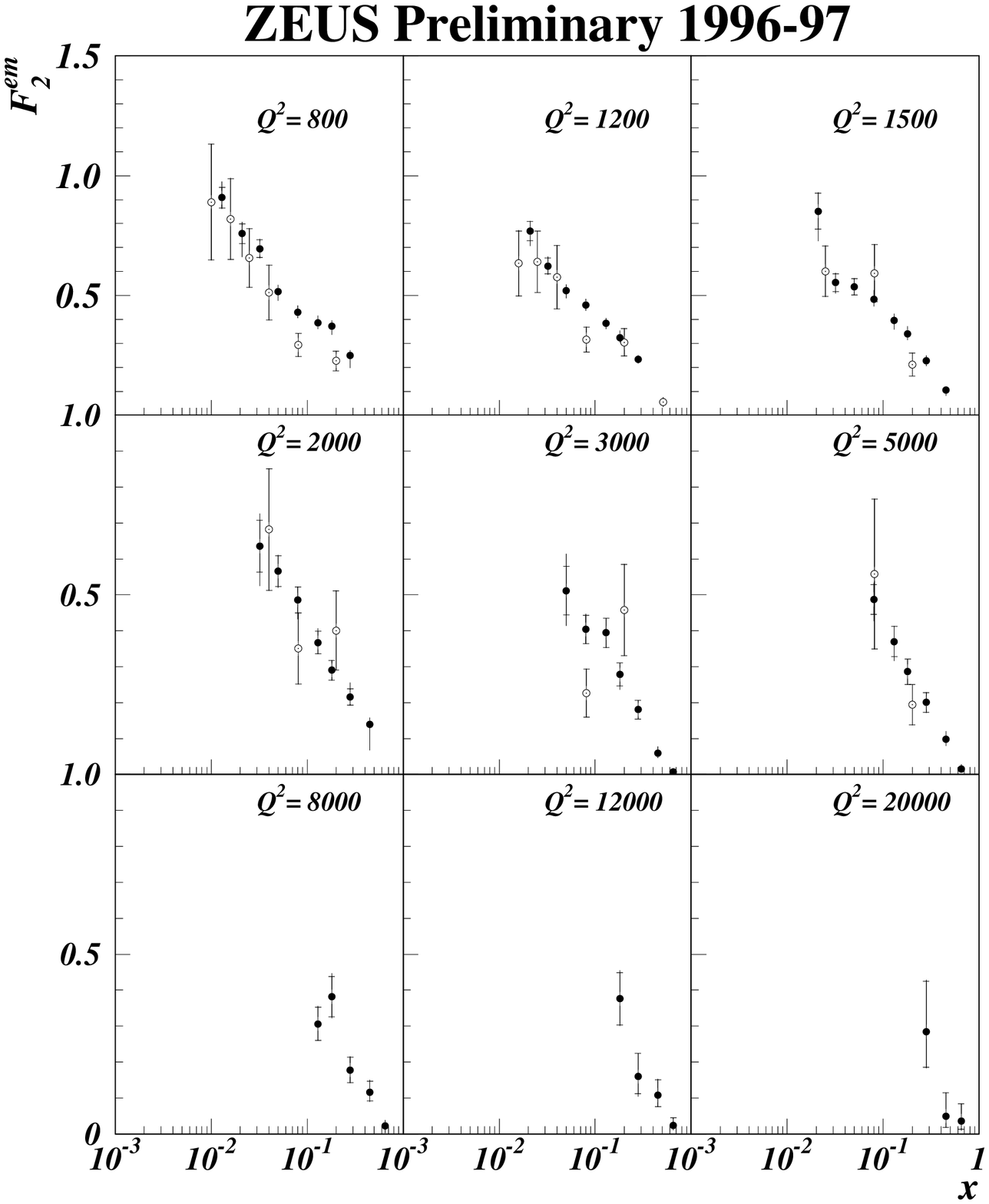}}
\vskip -.2 cm
\caption[]{
\label{figure9}
\small ZEUS 1996 1997 $F_2$ measurement at medium $Q^2$.}
\end{figure}

\begin{figure}[ht]      
\centerline{\epsfxsize 3.5 truein \epsfbox{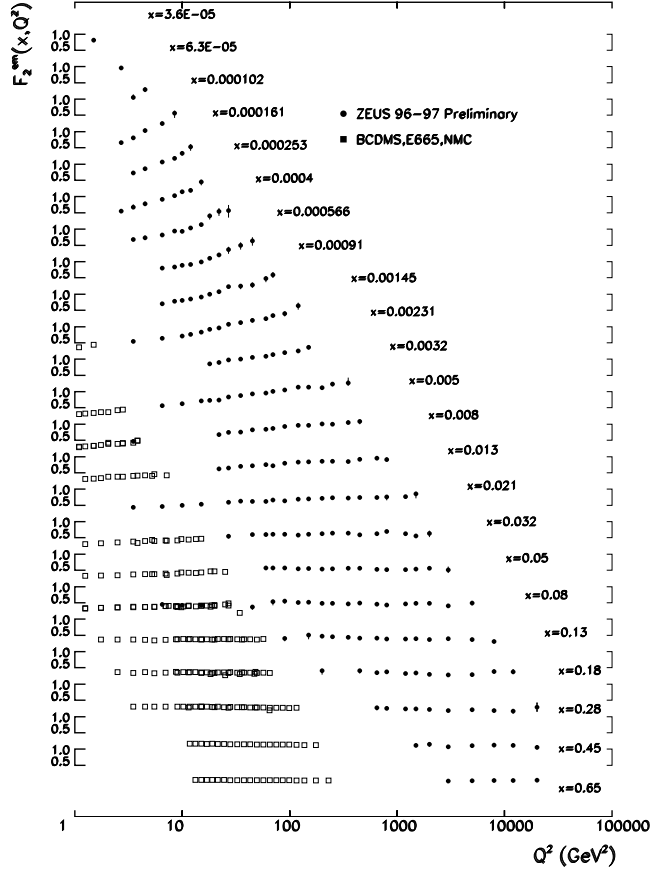}}
\vskip -.2 cm
\caption[]{
\label{figure10}
\small ZEUS 1996 1997 $F_2$ measurement for all $x$ and $Q^2$.  This plot 
highlights the rise in $F_2$ as $x$ decreases.}
\end{figure}

\end{document}